\documentclass[12pt,epsf]{article}
\usepackage{graphicx}
\usepackage{epsfig,amsmath,amssymb,verbatim,mathrsfs,subfigure,feynmf}
\begin{document}

\def\a{\alpha}
\def\b{\beta}
\def\c{\varepsilon}
\def\d{\delta}
\def\e{\epsilon}
\def\f{\phi}
\def\g{\gamma}
\def\h{\theta}
\def\k{\kappa}
\def\l{\lambda}
\def\m{\mu}
\def\n{\nu}
\def\p{\psi}
\def\q{\partial}
\def\r{\rho}
\def\s{\sigma}
\def\t{\tau}
\def\u{\upsilon}
\def\v{\varphi}
\def\w{\omega}
\def\x{\xi}
\def\y{\eta}
\def\z{\zeta}
\def\D{\Delta}
\def\G{\Gamma}
\def\H{\Theta}
\def\L{\Lambda}
\def\F{\Phi}
\def\P{\Psi}
\def\S{\Sigma}

\def\o{\over}
\def\beq{\begin{eqnarray}}
\def\eeq{\end{eqnarray}}
\newcommand{\gsim}{ \mathop{}_{\textstyle \sim}^{\textstyle >} }
\newcommand{\lsim}{ \mathop{}_{\textstyle \sim}^{\textstyle <} }
\newcommand{\vev}[1]{ \left\langle {#1} \right\rangle }
\newcommand{\bra}[1]{ \langle {#1} | }
\newcommand{\ket}[1]{ | {#1} \rangle }
\newcommand{\EV}{ {\rm eV} }
\newcommand{\KEV}{ {\rm keV} }
\newcommand{\MEV}{ {\rm MeV} }
\newcommand{\GEV}{ {\rm GeV} }
\newcommand{\TEV}{ {\rm TeV} }
\def\diag{\mathop{\rm diag}\nolimits}
\def\Spin{\mathop{\rm Spin}}
\def\SO{\mathop{\rm SO}}
\def\O{\mathop{\rm O}}
\def\SU{\mathop{\rm SU}}
\def\U{\mathop{\rm U}}
\def\Sp{\mathop{\rm Sp}}
\def\SL{\mathop{\rm SL}}
\def\tr{\mathop{\rm tr}}

\def\IJMP{Int.~J.~Mod.~Phys. }
\def\MPL{Mod.~Phys.~Lett. }
\def\NP{Nucl.~Phys. }
\def\PL{Phys.~Lett. }
\def\PR{Phys.~Rev. }
\def\PRL{Phys.~Rev.~Lett. }
\def\PTP{Prog.~Theor.~Phys. }
\def\ZP{Z.~Phys. }
\newcommand{\draftnote}[1]{\textbf{#1}}


\baselineskip 0.7cm

\begin{titlepage}

\begin{flushright}
IPMU11-0195
\end{flushright}

\vskip 1.35cm
\begin{center}
{\large \bf  Large Mixing Angles From Many Right-Handed Neutrinos}
\vskip 1.2cm

Brian Feldstein$^{a}$ and  William Klemm$^{a,b}$
\vskip 0.4cm

${}^{a}${\it  Institute for the Physics and Mathematics of the Universe, \\
University of Tokyo, Kashiwa, 277-8568, Japan}\\

${}^{b}${\it Department of Physics and Astronomy, \\
Uppsala University, Box 516, SE-751 20 Uppsala, Sweden}\\

\vskip 1.5cm

\abstract{
A beautiful understanding of the smallness of the neutrino masses may be obtained via the seesaw mechanism, whereby one takes advantage of the key qualitative distinction between the neutrinos and the other fermions: right-handed neutrinos are gauge singlets, and may therefore have large Majorana masses.  The standard seesaw mechanism, however, does not address the apparent lack of hierarchy in the neutrino masses compared to the quarks and charged leptons, nor the large leptonic mixing angles compared to the small angles of the CKM matrix.  In this paper, we will show that the singlet nature of the right-handed neutrinos may be taken advantage of in one further way in order to solve these remaining problems:  Unlike particles with gauge interactions, whose numbers are constrained by anomaly cancellation, the number of gauge singlet particles is essentially undetermined.  If large numbers of gauge singlet fermions are present at high energies -- as is suggested, for example, by various string constructions -- then the effective low energy neutrino mass matrix may be determined as a sum over many distinct Yukawa couplings, with the largest ones being the most important.  This can reduce hierarchy, and lead to large mixing angles. 
Assuming a statistical distribution of fundamental parameters, we will show that this scenario leads to a good fit to low energy phenomenology, with only a few qualitative assumptions guided by the known quark and lepton masses.  The scenario leads to predictions of a normal hierarchy for the neutrino masses, and a value for the $|m_{\rm {ee}}|$ mass matrix element of about $1-6$ meV. 
}

\end{center}
\end{titlepage}

\setcounter{page}{2}

\section{Introduction}

The quark and lepton masses and mixing angles provide one of the fundamental mysteries of the standard model of particle physics.  They hint at a potentially deep underlying structure, while also seeming sufficiently random so as to defy straightforward explanation.  There are a number of qualitative features of these parameters which one would like to understand:\footnote{The strong CP problem may or may not have a solution related to the other flavor mysteries, and we will not concern ourselves with it in this paper.  We may rely, for example, on a separate solution involving an invisible axion with a large Peccei-Quinn symmetry breaking scale.}
\begin{enumerate}
\item{Why are the neutrino masses roughly ten orders of magnitude smaller than the masses of the quarks and charged leptons?}
\item{Why do the quark and charged lepton masses have significant hierarchies of about five orders of magnitude?}
\item{Why is the CKM matrix approximately equal to the identity when the up and down quarks are both ordered by mass?}
\item{Why are the mixing angles in the lepton sector fundamentally different from those in the quark sector, with two angles close to maximal?}
\end{enumerate}

Of these questions, it is the first which has lent itself most easily to explanation.  The key observation is that right-handed neutrinos, unlike any of the other standard model fermions, have no known gauge interactions.  As a result, no symmetry forbids them from obtaining Majorana masses, which may be many orders of magnitude larger than the weak scale.  After integrating out the right-handed neutrinos from the theory, one obtains effective operators of the form $(HL)^2/M_R$, where $H$ is the Higgs field, $L$ is a lepton doublet, and $M_R$ is the right-handed neutrino mass scale.  This yields light Majorana neutrinos with masses of order $v^2/M_R$, with $v=174$ GeV being the Higgs vacuum expectation value, and with a possible additional suppression from Yukawa couplings \cite{seesaw}.  Appropriately small masses are then obtained for $M$ of order $10^{14}-10^{16}$ GeV, perhaps related to the scale of grand unification.  This picture is so simple that it is now taken almost for granted, although it should be kept in mind that experimental confirmation is still lacking.

The remaining questions on our list, on the other hand, have no such obvious interpretations, although a large variety of  proposals have certainly been put forth.   In many models, one assumes that the quarks and leptons are charged under a variety of possible flavor symmetries, with the Yukawa couplings proportional to appropriate powers of symmetry breaking spurions.  While such models can produce successful phenomenology, the choices for the flavor symmetry transformations, as well as the associated spurion structures, can seem ad hoc, and it is perhaps fair to say that no particular model stands out as being especially compelling.

Another possibility, which runs somewhat counter to the flavor symmetry perspective, is that the various Yukawa couplings are determined in some more complicated way by the fundamental theory, with their low energy values not representing any essential pattern.   In this scenario, one simply parametrizes the Yukawa couplings via a hypothesized statistical distribution, and checks the phenomenological consequences.  Such distributions may arise, for example, via the landscape of string theory, or may perhaps be manifestations of our own ignorance of the fundamental theory.
At the very least, the extent to which such an approach is viable is an important issue for determining the merit of the flavor symmetry paradigm. 
Fortunately, the standard model contains a large number of Yukawa couplings, and these may then serve as a reasonable statistical sample from which to draw proposals for the Yukawa distributions.  A key issue with this approach, however, comes from question number (4), above.  Naively, it seems difficult to allow for a single form of probability distribution to describe all of the standard model Yukawa couplings.  The quarks and charged leptons seem to prefer a probability distribution which scans roughly evenly over many orders of magnitude in order to yield large mass hierarchies and the small CKM mixing angles \cite{Donoghue}.  The neutrinos, on the other hand, would seem to prefer a more degenerate -- or ``anarchical" type of scan in order to obtain the required large mixing angles \cite{anarchy1, anarchy2, anarchy3}.  While we currently have no way to derive these probability distributions from fundamental theory, and it is certainly possible that the neutrino Yukawas are simply determined in a special way, different from the other fermions, a unified description would seem more attractive, and would perhaps lend more weight to this approach.

In this paper we will show that the seesaw mechanism alone may be what is distinguishing the neutrinos from the other fermions, both in terms of their masses, and also in terms of their large mixing angles.  In particular, we will show that it is indeed possible to make use of the same general form of distribution for all of the Yukawa couplings of the standard model, and still produce appropriate phenomenology.  We will consider scans with large hierarchies, which seem naively most suited to the quarks and charged leptons.  As in the standard seesaw, for the neutrinos we shall take advantage of the fundamental physical feature which distinguishes them from the other fermions; the fact that the right-handed neutrinos are gauge singlets.  As a result, not only do the right-handed neutrinos have the possibility for large Majorana masses, but moreover, due to the absence of any anomaly cancellation requirements, {\it we do not know how many of them there are}.  Indeed, constructions in string theory often produce large numbers of singlet fermions with Majorana masses close to the GUT scale after the compactification of extra dimensions \cite{Buchmuller, Vafa, Watari}.\footnote{For more general field theory discussions, see also \cite{Ellis, Schechter}.}

In the presence of a large number of singlets, the low energy neutrino mass matrix will be realized as a sum over many distinct hierarchical numbers with the largest Yukawa couplings dominating.  This generically results in a washed out hierarchy, and potentially large mixing angles.  It follows that a random hierarchical scan of Yukawa couplings is in fact a good fit for the observed masses and mixing angles of the standard model, so long as the possibility for a large number of right-handed neutrinos is taken into account.  

The reason that many singlet states often arise with masses close to the GUT scale in string constructions is actually reasonably generic:  After compactifying $d$ extra dimensions in a string model, the resulting Kaluza Klein mass scale is related to the effective four-dimensional Planck scale $M_{{\rm Pl}}$ through the relation
\begin{equation}
M_{\rm{KK}} \sim M_{{\rm Pl}} \left(\frac{M_{\rm{KK}}}{M_s}\right)^{\frac{d}{2} + 1},
\end{equation}
where $M_s$ is the string scale.  In order for the compactification geometry to be reasonably described by a classical gravity picture, it is required that $M_{\rm{KK}}$ be at least parametrically smaller than $M_s$.  From the above relation, and considering the 6 available extra dimensions of string theory, we thus see that it may be reasonable to expect the Kaluza Klein scale to be roughly a few orders of magnitude smaller than $M_{{\rm Pl}}$.\footnote{It is not necessarily required, however, that the compactification of the extra dimensions have a well defined classical gravity description.  This is perhaps the main caveat to the present argument.}  Note that it is also possible that grand unified symmetry is broken via a mechanism directly related to the compactified extra dimensions, further motivating the presence of the KK scale at close to $10^{14}-10^{16}$ GeV.  In any case, the point then is that the KK scale is a natural scale for the appearance of a large number of gauge singlets which may then serve as right-handed neutrinos;  these singlets might be born out of KK towers themselves, as was the case considered in \cite{Vafa}, or they might be associated with moduli fields, stabilized by masses close to $ M_{\rm{KK}}$ due to fluxes in the compactified dimensions \cite{Watari}.\footnote{In this latter case note that the singlet masses are actually given by $\sim M_{\rm{KK}}^3/ M_s^2$, with a further small suppression below the KK scale.}  For our purposes, the main point is that the number of relevant singlets with masses close to the needed seesaw scale may easily number in the tens or hundreds.  Note that we will not attempt to construct an explicit top-down model for Yukawa couplings in this paper, but will simply take a phenomenological point of view based on the known properties of the standard model masses and mixing angles.  We will leave top-down model building as a possible subject for future work. 

The outline of our paper is as follows:  In section 2 we will describe our framework in more detail, discussing the various qualitative features we require, including the properties of the high energy Majorana mass matrix.  In section 3, we will give an example of a particular universal Yukawa coupling distribution which gives a good phenomenological fit for the charged fermion masses and mixings, and show how our mechanism then also leads to good phenomenology in the neutrino sector.  Additionally, we will show that our setup leads to the predictions of a normal hierarchy for the neutrino masses, and a value of $|m_{\rm{ee}}|$ -- the neutrino mass parameter relevant for neutrinoless double beta decay -- of between about $1$ and $6$ meV.  We will summarize and discuss future directions in section 4.

\section{Framework}

The reason that many right-handed neutrinos, which we shall denote generically as $\nu_R's$, can wash out hierarchy in the low energy neutrino mass matrix $m$ is simple.  This matrix takes the form
\begin{equation}
m_{ij} = v^2 \sum_{lk} Y_{ik} M^{-1}_{kl} Y_{jl}, \label{sum}
\end{equation}
where, with $N$ right-handed neutrinos, $Y$ is the $3\times N$ Yukawa matrix, $M$ is the $N\times N$ Majorana mass matrix, and $v$ is the Higgs vacuum expectation value.  Taking the typical largest Yukawa coupling size  to be of order one, for example, we can see that, with sufficiently many $\nu_R$'s, each matrix element in $m$ will generically obtain some number of large contributions.  Even if the original Yukawas were hierarchical, the $m_{ij}$'s then come out to be roughly of the same order, with the hierarchy having been lost (note that, assuming arbitrary signs or phases for each term in the sum, one obtains an enhancement to the overall scale of $m$ by a factor which scales as $\sqrt{N}$).

The mechanism we are discussing here is fairly general, and a specific structure or distribution for the Yukawa couplings and Majorana masses is not necessary.  We do, however require the following conditions to be satisfied:
\begin{itemize}
\item There must be a reasonably well defined upper-bound to the neutrino Yukawa couplings.  Such an upper-bound could be set by perturbativity or by some other fundamental physics, and should be reasonably independent of which neutrino field a given Yukawa coupling is associated with.
\item Enough right-handed neutrinos must be present so that each left-handed neutrino is expected to have $\gsim 1$ Yukawa couplings within a factor of a few of the upper bound.  It can be seen by inspection of equation \ref{sum} that this would result in each low energy neutrino mass matrix element expecting to receive at least one ``large" contribution.  If the Majorana mass matrix has hierarchical eigenvalues, then this condition applies to the lightest set of the $\nu_R$'s, which will give the dominant contribution to the seesaw.  Amongst the standard model quarks and charged leptons, we have one order one Yukawa coupling amongst 27 (associated with the top quark), and so with this as a guide, we anticipate that we may require roughly $\gsim 30$ right handed-neutrinos for our mechanism to work.
\item We require a Majorana mass matrix which mixes together the right-handed neutrino fields.  One simple way this might be realized is if the physics determining the Majorana masses is distinct from that determining the Yukawa couplings (in which case the unitary matrix which diagonalizes the Majorana masses can be taken to be distributed according to the Haar measure, as will be discussed in section 3), but essentially any sufficiently non-diagonal structure should be viable.
\end{itemize} 
The need for this last condition can be seen as follows:  Suppose that the Majorana mass matrix were diagonal.  In that case there would be a tendency for the diagonal elements of the low energy mass matrix to be larger than the off-diagonal elements, resulting in small mixing angles.  Let us compare for example the $1,1$ element to the $1,2$ element:
\begin{equation}
m_{11} \sim \sum_{k} M^{-1}_{kk} Y_{1k}^2, \label{sum11}
\end{equation}
\begin{equation}
m_{12} \sim \sum_{k} M^{-1}_{kk} Y_{1k} Y_{2k}. \label{sum12}
\end{equation}
For any given Yukawa coupling $y$, there is in general a much higher chance that $y^2$ will be near its upper bound, than the chance that two separate couplings $y_1$ and $y_2$ will both be large and lead to a large product $y_1y_2$.  It is thus clear from equations \ref{sum11} and  \ref{sum12} that a diagonal Majorana mass matrix will tend to give poor phenomenology in our scenario.\footnote{Note that, even with a Majorana mass matrix with substantial off-diagonal components, squared Yukawa couplings will still only contribute to the diagonal elements of $m_{ij}$.  With many right-handed neutrinos, however, the large number of terms (of order $N^2$) which are products of separate Yukawa couplings actually give the dominant contribution.}

It should be clear that we certainly do not require that all of the Yukawa couplings of the quarks and leptons share a universal distribution.  We will, however, concentrate on such distributions in this work, since in this way we may check that it is actually the many $\nu_R$'s which are resulting in large mixing angles, rather than any fundamental lack of hierarchy assumed for the neutrino Yukawa couplings.  Note that for this reason, there is no intrinsic obstacle to realizing our scenario in a supersymmetric context (in which $\tan{\beta}$ would modify the hypothetical quark and charged lepton distributions somewhat), or even in the context of a grand unified theory. 

Note that a wave-function overlap picture (see e.g. \cite{Vafa, Watari, Salem1, Salem2}) for obtaining hierarchy in the Yukawa couplings would not work very well for our scenario, since in such cases the couplings for a given field tend to all be correlated in size.  This is the key manner in which our scenario with many right-handed neutrinos differs from those already appearing in the literature.  Constructions with compactified extra-dimensions may still be used to motivate the presence of the many $\nu_R$ states near the GUT scale, but we require, for example, extra-dimensional wave function profiles to be fairly flat, with the Yukawa hierarchy generated by fundamental physics in some other manner.

In the next section we will show our mechanism at work quantitatively in a specific example:  We will take a Yukawa coupling distribution fit to the known properties of the quarks and charged leptons, and work out the phenomenological consequences as a function of the number of $\nu_R$'s.  We will also show that our scenario (as well as any scenario with a roughly anarchical low energy neutrino mass matrix) leads to a prediction of $|m_{{\rm ee}}| \sim 1 - 6$ meV.

\section{An Example}

\subsubsection*{Fitting to the charged fermion masses and CKM angles}
Without knowledge of the high-scale theory it is of course impossible to know the fundamental distribution of the Yukawa couplings.  However, existing data can serve as a guide; the best fitting probability distribution for the charged fermions has been examined in some detail in \cite{Donoghue}, which we use as a starting point for our study.  For power law distributions with minimum and maximum cutoffs, they found that the quark and charged lepton masses are best fit by a distribution that is very close to scale invariant, $\rho(m)\sim 1/m$.  However, we expect that the Yukawa couplings, not the  masses, are the fundamental parameters of the underlying theory.  Because of mixing effects, the form of the Yukawa probability distribution is not directly transferred to the masses.  Considering distributions of the form 
\beq
\rho(y)\sim {1\over y^\delta} \label{eq:distribution}
\eeq
at the GUT scale within the standard model, the authors of \cite{Donoghue} found that values of $\delta$ just below $1.2$ give the optimal mass distribution, with $1.2\times 10^{-6}\leq y\leq 1$.

\begin{figure}[h]
\begin{center}
\includegraphics[width=10cm]{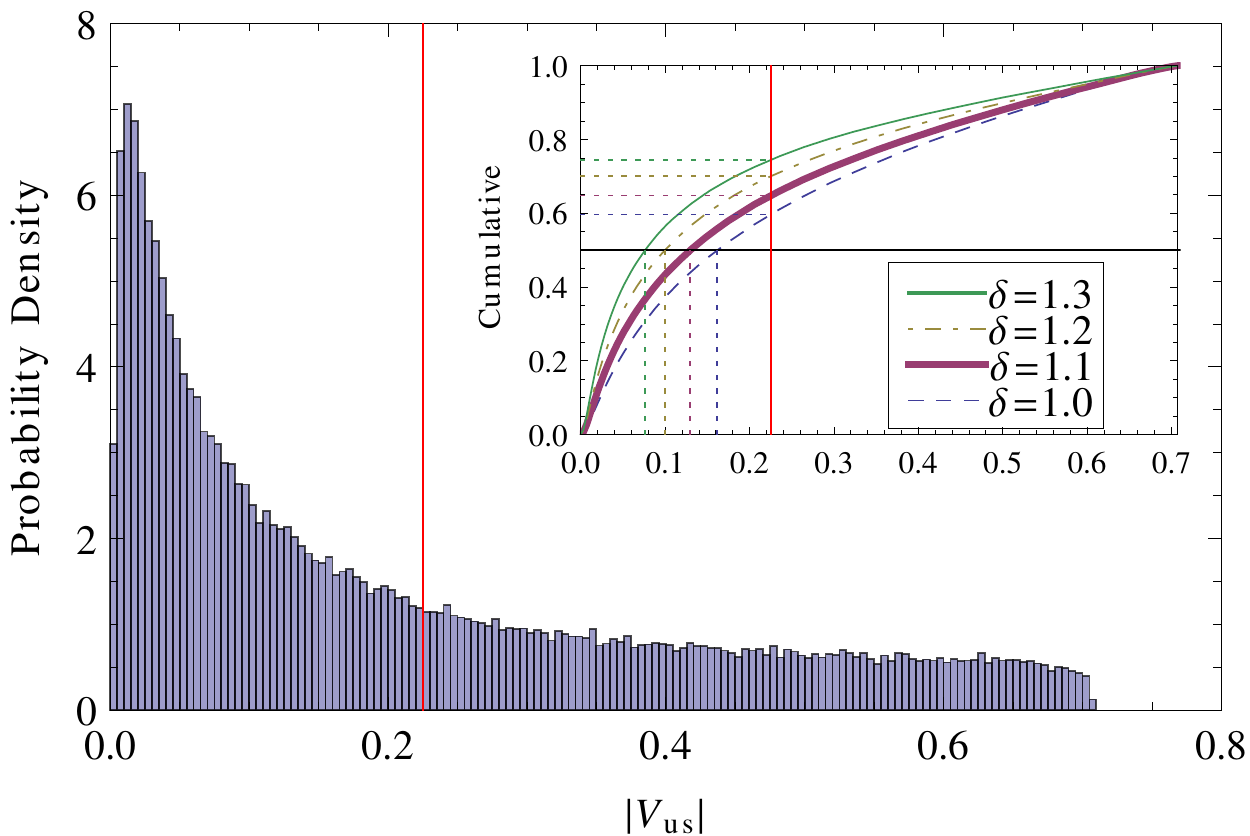}
\caption{$|V_{us}|$ for $\delta=1.1$ for matrices with mass eigenvalues within a factor of $\sqrt{10}$ of GUT scale values.  
The vertical red line represents the experimental value of $|V_{us}|$.  The maximum possible value is $1/\sqrt{2}$ due to our requirement that the diagonal entries be the largest. Inset: Cumulative probability distribution function for   
$\delta=1.0-1.3$. Vertical dotted lines indicate the median values of the distributions, and horizontal dotted lines indicate the fraction of matrices with $|V_{us}|$ less than the experimental value.}
\label{fig:Vus}
\end{center}
\end{figure}

The Yukawa couplings also determine fermion mixing, so we wish to ensure that our choice of distribution also leads to reasonable values for the CKM elements.  In fact, the hierarchical structure considered here naturally leads to small mixing angles.  Note however, that there is a caveat; without any flavor symmetry to correlate the Yukawa couplings of a given field, we must rely on a minor coincidence to ensure that it is actually the mass ordered basis in which the CKM matrix appears approximately diagonal.  The probability that this could have occurred by chance is $1/3! \sim 17\%$, and we will take this as a starting assumption, considering only CKM matrices in which the largest element of each row and column lies on the diagonal in the mass ordered basis.
For the distribution considered in \cite{Donoghue}, with $\rho(y)\sim 1/y^{1.16}$, while it was found that reasonable masses and small CKM mixing were obtained, in some cases the observed CKM off-diagonal elements were {\it larger} than the mean predicted values, lying on tails of their distributions.  This can be seen from the plot of the $|V_{us}|$ distribution in Figure \ref{fig:Vus}.

To ensure that we consider a distribution which gives reasonable quark mixings, we have generated a large number of up and down quark mass matrices, by scanning according to equation \ref{eq:distribution} for various values of $\delta$ and assigning each element a random phase. We then consider those which give a mass eigenvalue for each quark which lies within an order of magnitude of the measured value, i.e. in the range $( m_q/\sqrt{10}, m_q \sqrt{10})$.\footnote{Here and throughout this paper we use charged fermion mass eigenvalues evaluated at the GUT scale in the standard model \cite{hep-ph/0010004}, though \cite{Donoghue} demonstrated that the best fit for the distribution feels little effect from the running.  The experimental values for the CKM elements we cite do not reflect any running, but this effect is small, no more than $\sim 15\%$, and 
is negligible for $V_{us}$ \cite{Balzereit:1998id}.}
Finally, we diagonalize the mass matrices with eigenvalues ordered by mass and determine the corresponding CKM matrices, eliminating any with an incorrect generation structure, as discussed above.  As we see in Figure \ref{fig:Vus}, there is a sharp peak at small mixing, so to get a measure of how consistent each scenario is with the data, we characterize each distribution by its median as well as the probability of finding a value smaller than the experimentally observed value. We will use the 68\% and 95\% levels of the latter to define $1\sigma$ and $2\sigma$ ranges, as in \cite{Donoghue}. This is illustrated for $|V_{us}|$ in Figure \ref{fig:Vus} (inset), which shows some preference for smaller values of $\delta$ -- smaller $\delta$ leads to larger mixing angles.  To get an overall measure of the CKM matrix, we consider the three off-diagonal elements shown in Table \ref{tab:CKM}. The measured values of $|V_{us}|$, $|V_{ub}|$, and $|V_{cb}|$, all show improvement of agreement as $\delta$ is decreased, and fall within our $1\sigma$ criterion around $\delta=1.1$.  We use this, along with the mass only best fit of $\delta\sim 1.2 \pm .1$ from \cite{Donoghue}, in our choice of distribution.  While one could perform a detailed statistical analysis to determine the best fit to all measured parameters, the $\delta=1.1$ case gives rise to both reasonable charged fermion masses as well as CKM angles, so we will take it as our example distribution for studying the neutrino sector.

\begin{table}
\begin{center}
\begin{tabular}{|c||c|c|c||c|c|c|}
\hline
{} &\multicolumn{3}{|c||}{Median} & \multicolumn{3}{c|}{\% below exp.}\\\hline
$\delta$ & $V_{us}$& $V_{ub}$& $V_{cb}$& $V_{us}$& $V_{ub}$& $V_{cb}$\\\hline
1 & .162 & .018 & .060 & 60 & 19 & 42\\\hline
1.1 & .130 & .012 & .032 & 65 & 24 & 52\\\hline
1.2 & .099 & .007 & .014 & 70 & 32 & 63\\\hline
1.3 & .076 & .004 & .006 & 75 & 40 & 73\\\hline\cline{1-4}
{\bf Exp.} & \bf .2252 &\bf .00389 &\bf .0406 & \multicolumn{3}{c}{}\\\cline{1-4}
\end{tabular}
\caption{Median values of CKM element magnitudes for different values of $\delta$ along with their measured values \cite{pdg}.  The rightmost columns indicate the percent of matrices which give a given CKM element smaller in magnitude than the measured value. Restrictions on the mass eigenvalues are as in Figure \ref{fig:Vus}.}
\label{tab:CKM}
\end{center}
\end{table}

\subsubsection*{The Majorana distribution}
For the high energy Majorana mass matrix, for which we have no guidance from data, we take a simple ansatz in accord with the discussion of section 2; we suppose that the physics determining the Majorana mass matrix is independent from that determining the Yukawa couplings.  This assumption then implies that the unitary matrix $U$ which diagonalizes $M=U D U^T$ may be taken as distributed according to the Haar measure \cite{anarchy2}.\footnote{The Haar measure $\mu$ on a group $G$ satisfies $\mu(gE)=\mu(E)$ for every $g\in G$, and is uniquely defined for $G=U(N)$.} Haar distributed unitary matrices are easily generated 
via previously established algorithms \cite{generatehaar}.  Note that  we must further impose a distribution on the mass eigenvalues appearing in the diagonal matrix $D$; for simplicity we choose the case of a linear distribution between zero and $M_{R}$, the Majorana mass scale, which we fix by requiring that we match the value for the larger measured neutrino mass difference, $\Delta m_{31}^2$.

\subsubsection*{Probability for large PMNS mixing angles}
Given our chosen statistical distributions, we next
generate a large number of PMNS matrices for various numbers of right-handed neutrino species between $N=2$ and $N=100$.  We first generate a $3\times N$ Dirac mass matrix 
$M_D={v} Y$, where the entries of the Yukawa matrix $Y$ scan between $1.2\times 10^{-6}$ and 1 with probability density $\rho(y)\sim y^{-1.1}$ and have random phases.  We then generate a Haar distributed $N\times N$ Majorana mass matrix with linearly scanning eigenvalues, to form a neutrino mass matrix $m\equiv M_D M^{-1} M_D^T = U_\nu D_\nu U_\nu^T $, where $D_\nu$ is a diagonal matrix containing the neutrino mass eigenvalues, and $U_{\nu}$ is unitary.  We generate $3\times 3$ charged lepton matrices $M_l$ with the same distribution as the Dirac matrices for neutrinos. To more appropriately model the true charged lepton mass matrix, we only take matrices which agree to within a factor of $\sqrt{10}$ of the GUT scale charged lepton masses, as described previously for the quarks.
We diagonalize $M_l = U_l D_l V_l^{\dagger}$, from which we can generate the PMNS matrix $U_{PMNS}=U_l^\dagger U_\nu$.  This can be parameterized, with unphysical phases rotated away, in the standard fashion with three mixing angles ($\theta_{ij}$) and three $CP$ phases ($\delta,\alpha_{21},\alpha_{31}$),
\beq
U_{PMNS}&=&\left(
\begin{array}{ccc}
c_{12}c_{13}& s_{12}c_{13}&s_{13} e^{-i\delta}\\
-s_{12}c_{23}-c_{12}s_{23}s_{13}e^{i\delta}&c_{12}c_{23}-s_{12}s_{23}s_{13}e^{i\delta}&s_{23}c_{13}\\
s_{12}s_{23}-c_{12}c_{23}s_{13}e^{i\delta}&-c_{12}s_{23}-s_{12}c_{23}s_{13}e^{i\delta}&c_{23}c_{13}
\end{array}
\right)\\&{}&\times{\rm~diag}(1,e^{i{\alpha_{21}\over 2}},e^{i{\alpha_{31}\over 2}}),
\eeq
where $s_{ij}=\sin(\theta_{ij})$ and $c_{ij}=\sin(\theta_{ij})$.\footnote{Note that in this paper we ignore running effects due to the neutrino Yukawa couplings.  Even at and above the seesaw scale these are expected to be small for the distributions we consider, since we take all of our Yukawa couplings to be smaller than 1, with only a handful approaching this upper bound.  If the bound on the Yukawas were taken larger, then renormalization group running could have important, albeit model dependent effects.  For example, with low energy supersymmetry, depending on the superpartner spectrum, it is possible for running at the seesaw scale to contribute to slepton mass splittings and flavor changing neutral currents \cite{Ellis}.  If only the standard model were assumed valid up to energies above the seesaw scale, then it is possible to obtain large radiative corrections to the Higgs quartic coupling and thereby large, experimentally excluded, values for the Higgs mass.  Note that this latter conclusion would be avoided in the presence of high scale supersymmetry.}  Experimentally, there are two large mixing angles, with a best fit (assuming a normal hierarchy) of $\sin^2(\theta_{12})_{BF}=0.312^{+0.017}_{-0.015}$ and maximal mixing for $\sin^2(\theta_{23})_{BF}=0.52^{+0.06}_{-0.07}$ ($1\sigma$)\cite{Schwetz:2011qt}.  The third mixing angle is known to be small, with a best fit of $\sin^2(\theta_{13})_{BF}=0.013^{+0.007}_{-0.005}$ at $1\sigma$ \cite{Schwetz:2011qt},
and there is growing evidence for a nonzero mixing \cite{Khabibullin:2011dt,doublechooz}.

\begin{figure}[h]
\begin{center}
\includegraphics[width=10cm]{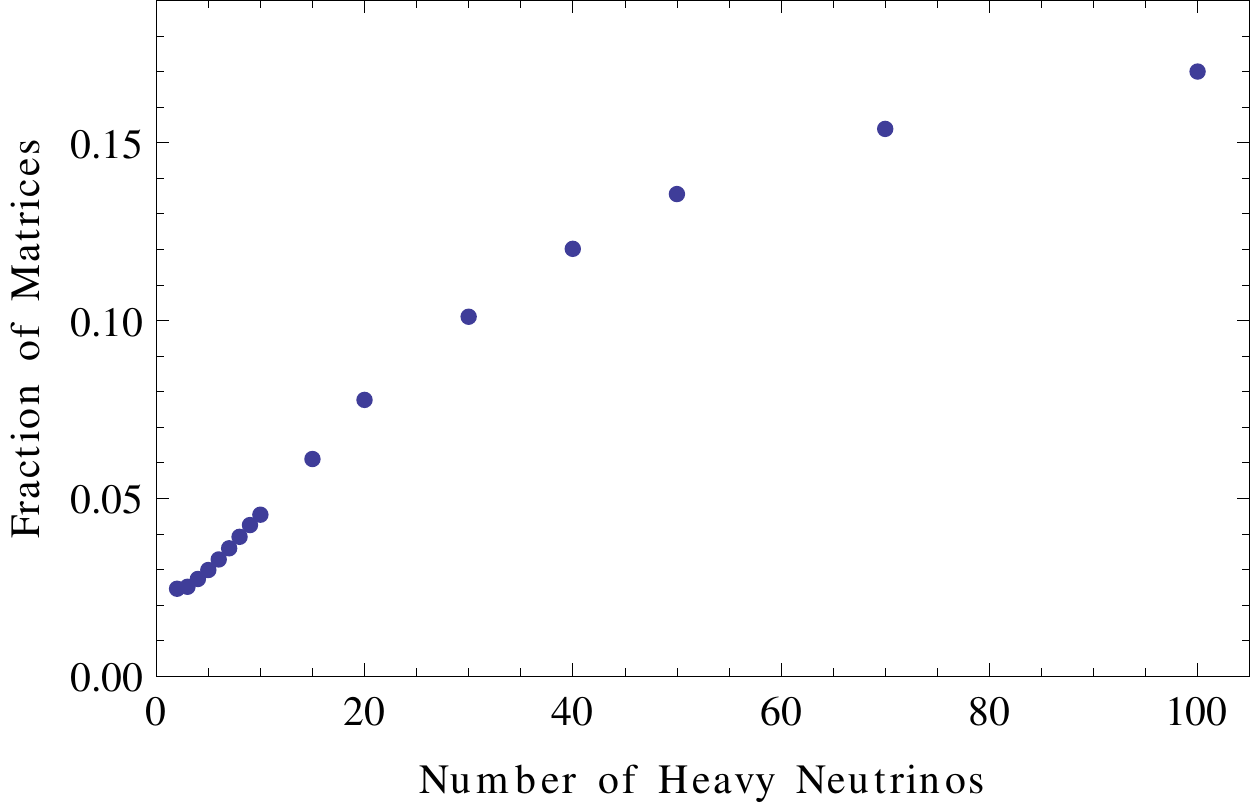}
\caption{Fraction of matrices with $\sin^2(2\theta)_{\rm largest}\geq 0.98$ and $\sin^2(2\theta)_{\rm next~largest}\geq 0.84$.}
\label{fig:2LMA}
\end{center}
\end{figure}

\begin{figure}[h]
\begin{center}
\includegraphics[width=10cm]{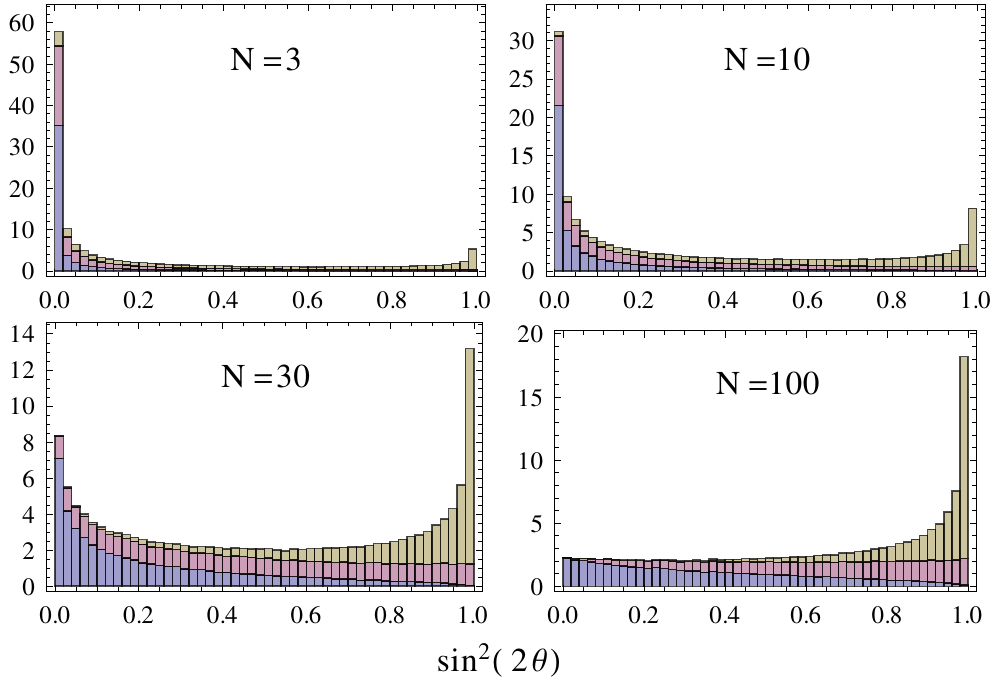}
\caption{Distribution of mixing angles.  The three different bands represent the largest, middle, and smallest $\sin^2(2\theta)$.}
\label{fig:angledist}
\end{center}
\end{figure}

To see how common such large mixing angles are in our scenario, for various numbers of right-handed neutrinos, we determine the fraction of matrices having both one near maximal mixing angle and one large mixing angle.  Because maximal mixing, with $\sin^2(2\theta)=1$, is a special point, we look for cases which have at least as much mixing as the $1\sigma$ experimental bounds, requiring that one angle satisfies $\sin^2(2\theta)\geq 0.98$ and another satisfies $\sin^2(2\theta)\geq 0.84$.  
The results are shown in Figure \ref{fig:2LMA}, from which we see a clear indication that as the number of right-handed neutrinos increases, so too does the likelihood of obtaining large mixing angles -- as expected for the reasons laid out in Section 2.
This effect is further illustrated in Figure \ref{fig:angledist}, where we see the shift to larger mixing angles as $N$ increases.

\subsubsection*{Other parameters}
\begin{figure}[h]
\begin{center}
\includegraphics[width=10cm]{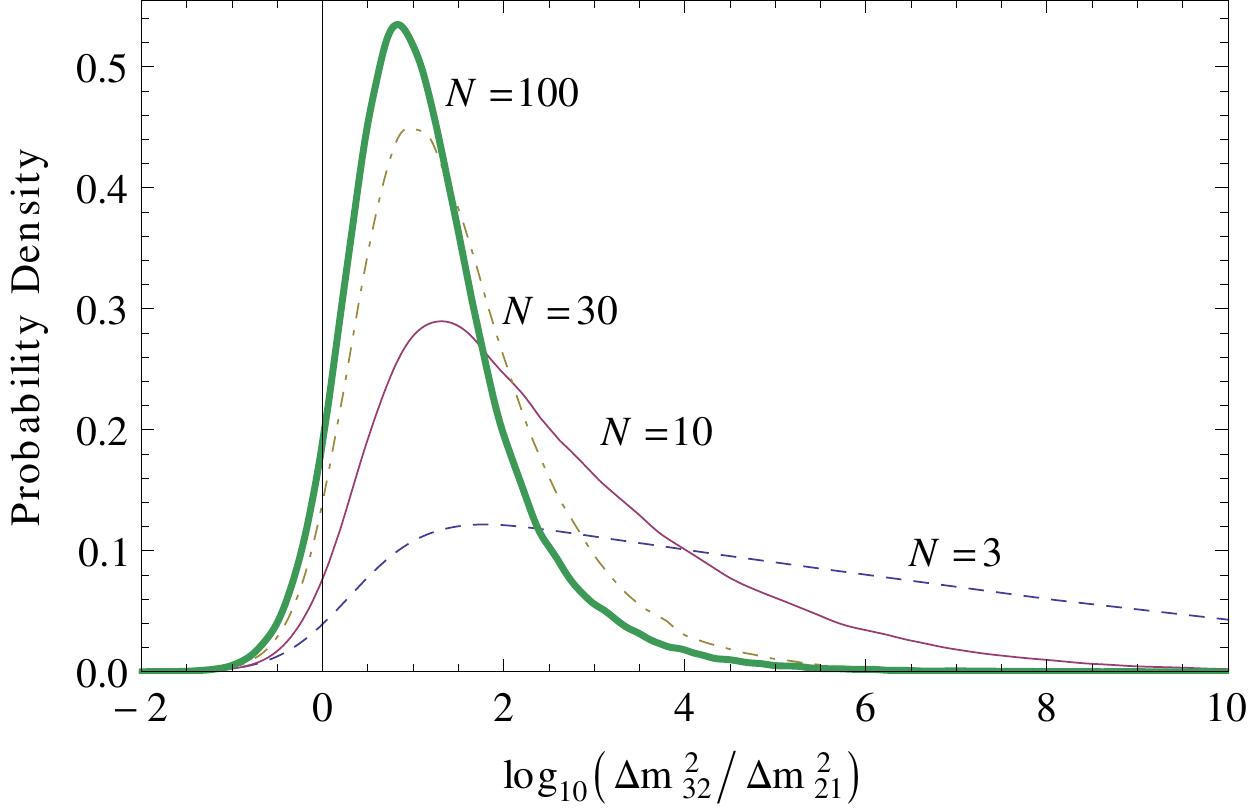}
\caption{Ratio of mass squared differences $\log_{10}\Delta m_{32}^2/{\Delta m_{21}^2}$ for $N=$ 3, 10, 30 and 100. Here we choose the convention $m_3>m_2>m_1$, so that positive(negative) values correspond to a normal (inverted) hierarchy.}
\label{fig:hierarchy}
\end{center}
\end{figure}

While the absolute masses of the neutrinos are not well measured, oscillation experiments give us a good measure of their mass squared differences, with a best fit of $\Delta m_{21}^2=7.59^{+0.20}_{-0.18}\times 10^{-5}{\rm~eV}^2$ and $\Delta m_{31}^2=2.50^{+0.09}_{-0.16}\times 10^{-3}{\rm~eV}^2$ (assuming a normal hierarchy, with comparable values for an inverted hierarchy) \cite{Schwetz:2011qt}.  To see if our construction accommodates this small but non-trivial hierarchy, and to determine whether there is a preference for a normal or inverted structure, in Figure \ref{fig:hierarchy} we consider the ratio of
neutrino mass squared differences, which we plot as $\log_{10}\Delta m_{32}^2/{\Delta m_{21}^2}$. Here we label the masses such that $m_3>m_2>m_1$, so that this quantity is positive for a normal hierarchy and negative for an inverted one.\footnote{Note that for an inverted hierarchy, our labeling is non-standard.}  Observed masses give a value of about $\pm 1.5$.  We see that for large $N$, the masses are much less hierarchical, and easily accommodate the observed values.  Furthermore, we see an overwhelming preference for the normal hierarchy, which in particular justifies our use of the associated mass and mixing angle measurements in later parts of this section.\footnote{
The reason our scenario strongly prefers a normal versus an inverted hierarchy is that the reasonably large observed ratio of solar and atmospheric mass squared differences necessitates that either the heaviest (normal hierarchy) or the lightest (inverted hierarchy) of the neutrinos is a mild outlier.  Having the heaviest neutrino as the outlier in our scenario is much more probable, since this requires fewer outlying elements in our typically degenerate mass matrix.}

\begin{figure}[h]
\begin{center}
\includegraphics[width=7.8cm]{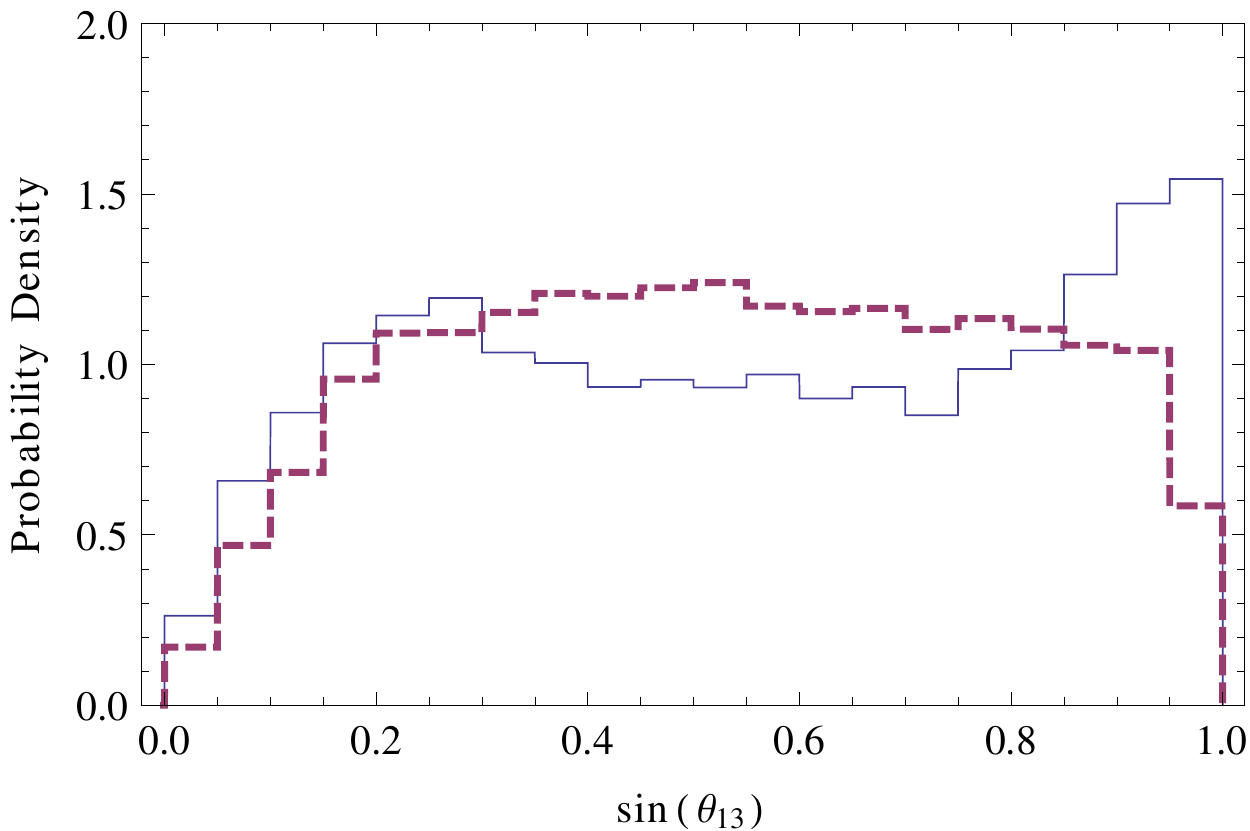}
\includegraphics[width=7.8cm]{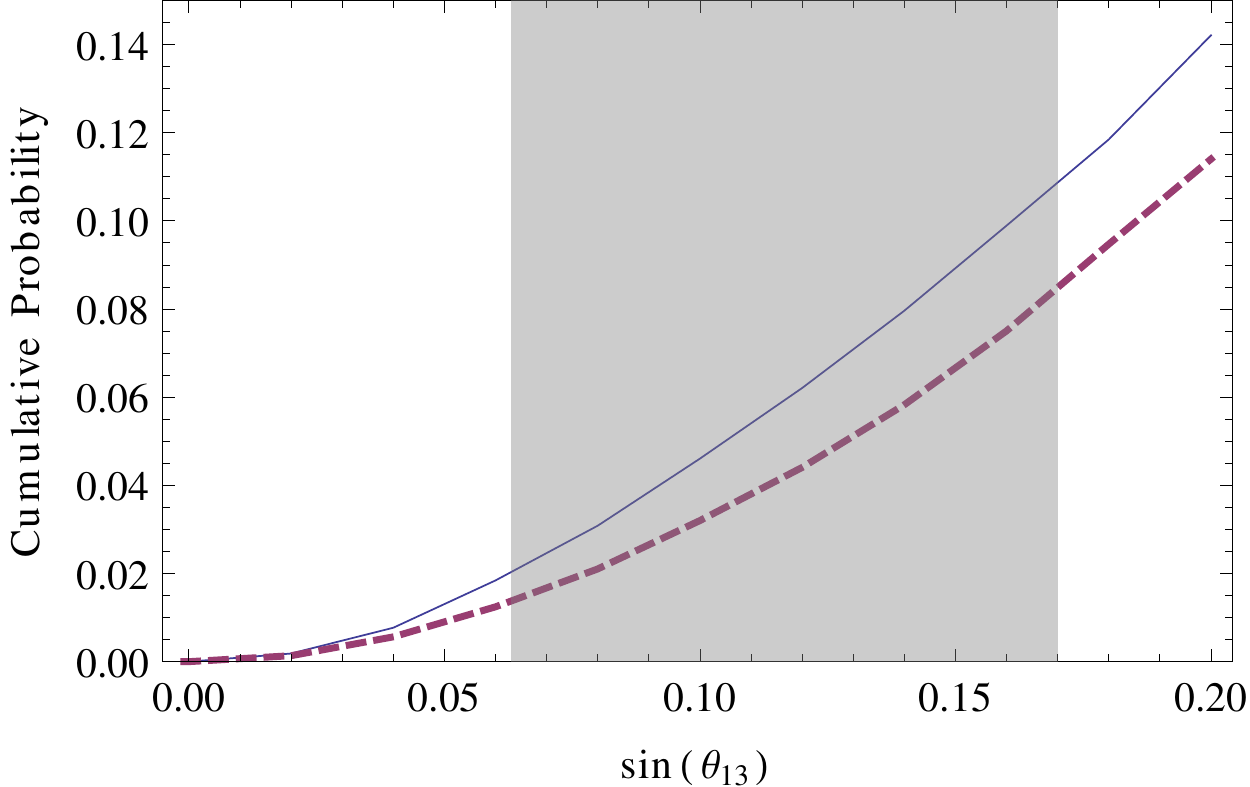}
\caption{Left: $\sin(\theta_{13})$ distributions among matrices satisfying $2\sigma$ large angle and mass constraints described in the text, for $N=30$ (blue, solid) and $N=100$ (red, dashed). Right: The corresponding cumulative probability distribution functions.  The grey shaded area indicates the $2\sigma$ best fit region. }
\label{fig:sintheta13}
\end{center}
\end{figure}

Having seen that the mixing angles and mass splittings observed in nature are increasingly typical as $N$ increases, we wish to look at other properties of viable matrices produced within our framework.  To select cases close to reality, we consider only matrices which satisfy: $0.28\leq \sin^2(\theta_{12})\leq 0.35$; $0.41\leq\sin^2(\theta_{23})\leq 0.61$; $29.1\leq \Delta m_{31}^2/\Delta m_{21}^2\leq 35.6$; and $0.004\leq\sin^2(\theta_{13})\leq 0.028$, which come from best fit $2\sigma$ bounds \cite{Schwetz:2011qt}.
In Figure \ref{fig:sintheta13}, we show the distribution of $\sin(\theta_{13})$, subject to the large angle and mass constraints, and find that there is some tension with the best fit, which at $2\sigma$ corresponds to about $0.06\leq \sin (\theta_{13})\leq 0.17$.  On the other hand, after fitting successfully the other parameters of the neutrino mass matrix, the probability of obtaining one mild outlier is not too small. This can be seen on the right side of Figure \ref{fig:sintheta13}, which gives the fraction of matrices at or below a given value of $\sin(\theta_{13})$ - there is significant variation across the experimentally preferred region (in grey), reaching around 10\% near the upper $2\sigma$ bound. Our scenario clearly prefers a nonzero value for $\sin(\theta_{13})$, for which there is growing evidence, and a significant range of allowed values would be unexceptional.
The situation for $\theta_{13}$ in our scenario is ultimately similar to that of neutrino anarchy, for which a global statistical analysis was performed and found good agreement with the data \cite{anarchy3}.  Performing a similar analysis for the present case is beyond the scope of this work, but we would expect to find similar results.

\begin{figure}[h]
\begin{center}
\includegraphics[width=10cm]{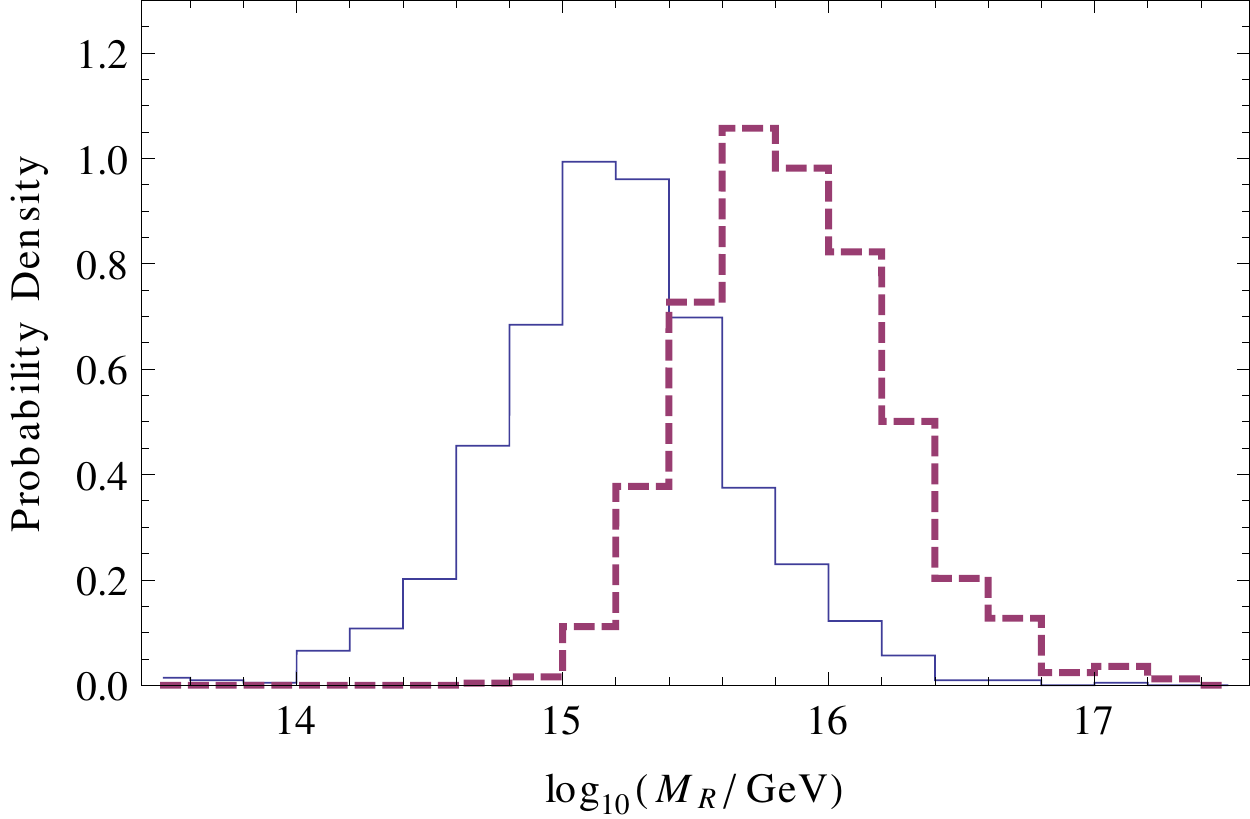}
\caption{Majorana mass scale distribution among matrices satisfying $2\sigma$ angle and mass constraints described in the text, for $N=30$ (blue, solid) and $N=100$ (red, dashed).}
\label{fig:Majorana}
\end{center}
\end{figure}

We next consider the Majorana mass scale, which we determine by requiring that the mass splitting $\Delta m_{31}^2$ matches its experimental value.  We plot the resulting distribution in Figure \ref{fig:Majorana}, where we see that the distribution rises and sharpens at large $N$, with typical values around $10^{15}-10^{16}{\rm~GeV}$, suggestive of the framework discussed in Section 2.

\begin{figure}[h]
\begin{center}
\includegraphics[width=10cm]{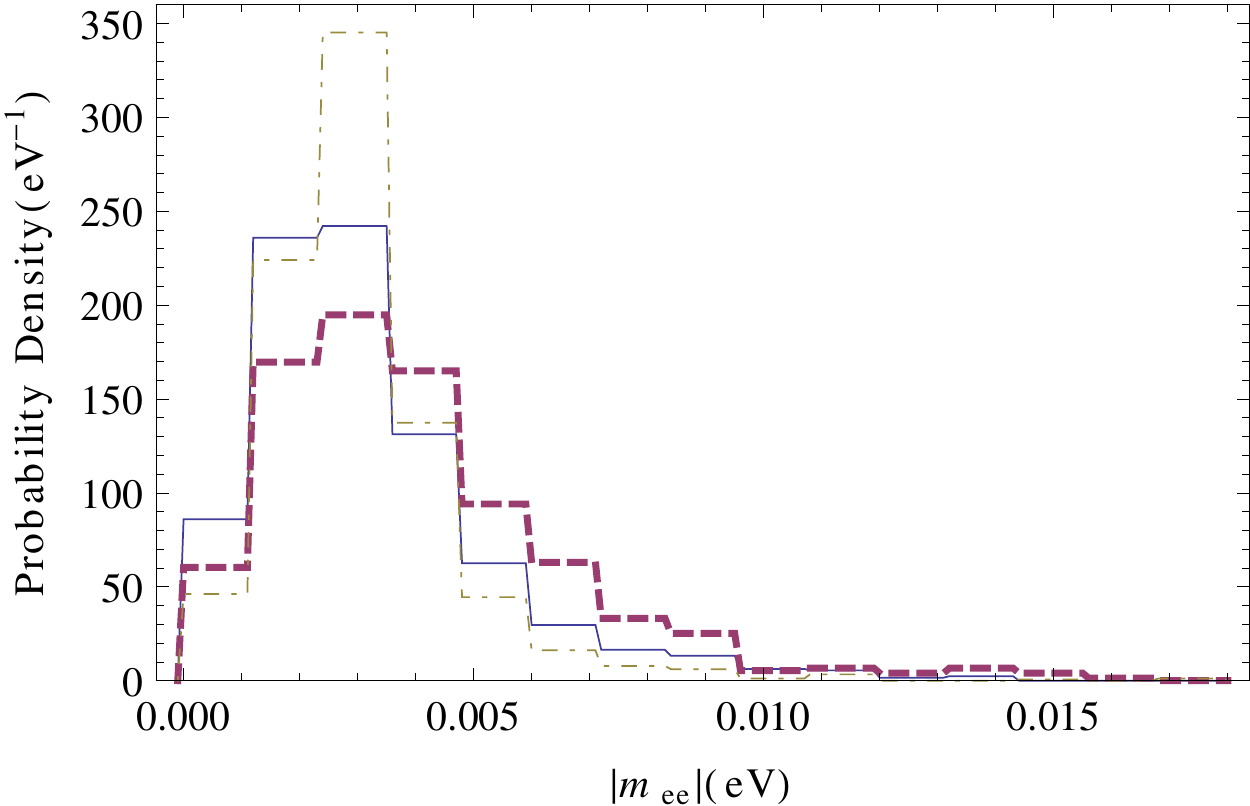}
\caption{$|m_{\rm{ee}}|$ distribution among matrices satisfying $2\sigma$ angle and mass constraints described in the text, for $N=30$ (blue, solid) and $N=100$ (red, dashed). Also shown is the anarchical case (yellow, dot-dashed) under the same constraints.}
\label{fig:mee}
\end{center}
\end{figure}

Neutrinoless double beta decay provides us with an experimental probe of Majorana interactions of neutrinos, and the figure of merit for such experiments is given by $m_{\rm{ee}}$, the upper left entry of the neutrino mass matrix in the diagonal charged lepton basis,
\beq
m_{\rm{ee}}=\sum_i U_{{\rm e}i}^2 m_i,
\eeq
where $U$ is the PMNS matrix and $m_i$ are the masses of the light neutrinos.  We show the prediction for $m_{\rm{ee}}$ in Figure \ref{fig:mee}, again using only matrices lying sufficiently close to observation, and with the overall scale set by the true value of $\Delta m_{31}^2$.  The $68\%$ central regions of the distributions are $.0015-.0049{\rm~eV}$ for $N=30$ and $.0018-.0062{\rm~eV}$ for $N=100$,
which seems typical for theories in which the low energy neutrino mass matrix is composed of relatively degenerate numbers -- for comparison, we also show the distribution we get by applying the same procedure to matrices with random ${\mathcal O}(1)$ parameters, i.e. anarchical matrices.\footnote{Here we generate anarchical matrices with basis-independent scanning following the prescription of \cite{anarchy2}: We generate a complex $3\times 3$ Dirac matrix, $M_D$, and complex symmetric $3\times 3$ Majorana matrix, $M$, each with entries scanning between 0 and 1 with random phases, subject to the basis-independent boundaries ${\rm Tr}(M^\dagger M$), ${\rm Tr}(M_D^\dagger M_D) \leq 1$. We then form our light neutrino mass matrix, $m=\Lambda M_D M^{-1} M_D^T$, where $\Lambda$ is chosen to obtain the correct value of $\Delta m^2_{31}$.  Because of the basis independence, we are free to choose the charged lepton matrix to be diagonal.} 
Unfortunately, the predicted values for $m_{\rm ee}$ are beyond the reach of current experiments.  On a 10 year or longer timescale, it is possible that neutrinoless double beta decay searches, such as a future iteration of EXO \cite{802228}, or observations of large scale structure \cite{astro-ph/0603494,arXiv:0805.1920} may reach the required sensitivity.  On shorter timescales, it is interesting to note that our framework could be falsified if neutrinoless double beta decay were to be observed with a large measured value of $m_{\rm ee}$.

Finally, we have done a similar analysis for the distributions of CP phases; we have found that their distributions are relatively uniform, not providing us with any particular prediction.

\subsubsection*{Other distributions}
To get a sense of how much our results depend on our particular choice of probability distribution, in Table \ref{tab:otherdist} we look at the frequency of obtaining two large angles for a few different scenarios, which we take for the purpose of illustration rather than any particular physical motivation.  We again require one angle with $\sin^2(2\theta)\geq 0.98$ and another with $\sin^2(2\theta)\geq 0.84$. Table \ref{tab:otherdist} also lists the percentage of $N=100$ matrices which satisfy $0.004\leq\sin^2(\theta_{13})\leq 0.028$ ($2\sigma$ bound), after having first satisfied the $2\sigma$ bounds on $\sin^2(\theta_{12})$, $\sin^2(\theta_{23})$, and the mass splitting ratio, described above.  
Case I is the distribution ($\delta=1.1$) we have already considered in detail, and is listed for reference.  

\begin{table}
\begin{center}
\begin{tabular}{|c|c||c|c|c|c||c||c|}
\hline
{} & $\delta$ & $N=3$ & $N=30$ & $N=100$ & $N=200$ & $P_{100}(\theta_{13})$ & Comments\\\hline
I & 1.1 &  2.5& 10.1 & 17.0 & 19.0 & 6 &\\\hline
II & 1.3 &  2.5 & 3.2  & 6.7 & 10.8 & 11 &\\\hline
III & 0.9 &  5.3 & 17.6 & 20.2 &  20.9 & 5 &\\\hline
IV & 0.9 &    3.7 & 16.3  & 19.7 & 20.5 & 5 & No minimum cutoff\\\hline
V & 1.1 &  3.1 & 6.8 & 15.4 & 18.4 & 7 & Hierarchical Majorana\\\hline
\end{tabular}
\caption{Percent of events with $\sin^2(2\theta)_{\rm largest}\geq 0.98$ and $\sin^2(2\theta)_{\rm next~largest}\geq 0.84$ for different distributions and numbers of right-handed neutrinos, as described in the text.  Also shown (denoted $P_{100}(\theta_{13})$): for matrices satisfying $2\sigma$ constraints for $\theta_{12}$, $\theta_{23}$, and the squared mass difference ratio, the percentage which also satisfy the $2\sigma$ bounds on $\theta_{13}$, with $N=100$. 
}
\label{tab:otherdist}
\end{center}
\end{table}

Case II demonstrates the importance of the second condition listed in Section 2-- that enough right-handed neutrinos must be present so that each left-handed neutrino is expected to have  $\gsim 1$ Yukawa couplings within a factor of a few of the upper bound.  Here, the distribution steeply favors the low end of the spectrum ($\delta=1.3$), which makes sampling of the upper cutoff much rarer; as a result, many more right-handed neutrinos are needed to wash out the hierarchy.  Whereas there is a clear enhancement by $N=30$ for $\delta=1.1$, the rise in likelihood of large angles does not occur until much larger $N$ in this case. 
Because the steeper distribution favors smaller mixing angles for a given number of right-handed neutrinos, this case performs somewhat better than the others for the $\theta_{13}$ distribution.

In case III, we see a distribution logarithmically skewed towards larger couplings ($\delta=0.9$).  This makes our large $N$ effect much more dramatic, becoming noticeable for smaller values of $N$.  Additionally, because this distribution is well behaved all the way to zero, it is not necessary to impose a lower bound on the Yukawa couplings for this value of $\delta$, so we remove this requirement for case IV.  Because only values close to the top of the distribution are important at large $N$, this only has the small impact of removing a bit of probability from the top.  The change is more apparent at small values of $N$, where the lower parts of the distribution can be relevant in viable matrices.

Finally, we turn to the choice of distribution for the Majorana mass matrix, which we have previously been taking to be anarchical and basis independent.  In case V we take a very different approach, choosing the same hierarchical distribution ($\delta=1.1$) and cutoff structure for both Dirac and Majorana mass matrix elements. While this appears to suppress washout effects somewhat, the impact is fairly weak, as the hierarchy is largely lost in the inversion of the Majorana matrix; only a small relative suppression of probability remains at $N = 100$.

From these scenarios we see that while the quantitative behavior varies somewhat, the qualitative effect of increased probability of large mixing angles from many right-handed neutrinos applies broadly in the class of distributions we have laid out.  Furthermore, we see that while there is some tension with a small value of $\sin(\theta_{13})$, all cases have a significant amount of probability lying in the experimentally favored region.


\section{Discussions and Conclusions}

In this work we have shown that the origin of the unique flavor properties of the neutrinos -- small masses, and large mixing angles -- may
both have their origin in the seesaw mechanism.  Even if the neutrino Yukawa couplings have large hierarchies, as are observed for the charged fermions, the hierarchical structure may generically become washed out if the seesaw involves a large number of right handed neutrinos, numbering perhaps in the tens or hundreds.  We have given an explicit example showing the mechanism at work, with a statistical distribution of Yukawa couplings of a roughly scale invariant form, fit to the properties of the quarks and charged leptons.  After integrating out the right-handed neutrinos, the probability of obtaining a mixing angle as large as the observed near-maximal value of $\theta_{23}$,
along with a second angle as large as the observed value of  $\theta_{12}$ was found to be about $20\%$.  The mechanism is fairly general, and may work with a variety of Yukawa structures, so long as various conditions are satisfied as described in the text.  General predictions of the framework are a normal hierarchy for neutrino masses, and a neutrino mass matrix element $|m_{\rm{ee}}|$ of about $1-6$ meV.

There are a variety of possible directions for future work.  The first and most obvious one would be to try to build a top down model yielding hierarchical Yukawa couplings of an appropriate nature for our mechanism.  This is non-trivial, since the most common approaches to producing hierarchical Yukawa couplings -- flavor symmetries and extra dimensional wavefunctions -- yield correlations across all of the couplings of a given field which do not lend themselves well to our framework.  It may also be interesting to consider the issue of leptogenesis \cite{lepto1, leptogenesis} in our scenario.\footnote{For some previous work on leptogenesis with many right handed neutrinos, see \cite{manylepto}.}  If the largest neutrino Yukawa couplings are of order one, as in the case of the quarks, then the seesaw scale we require is fairly large -- of order $10^{15}-10^{16}$ GeV.  It then follows that washout processes will destroy any lepton asymmetry produced by the decays of the right-handed neutrinos.  As a result, it may be necessary to move away from the thermal leptogenesis paradigm, and construct a model involving, for example, out of equilibrium inflaton decays to right-handed neutrinos at temperatures much below the seesaw scale \cite{nonthermal, leptogenesis}.  One other possible future direction would be to consider the impact of anthropic selection effects on the types of Yukawa coupling distributions that we have been working with.  One amusing possibility, somewhat orthogonal to the direction we have been pursuing here, is that the fundamental Yukawa distributions for the charged fermions may actually be reasonably degenerate, but with strong anthropic selection effects constraining the first generation quarks and charged leptons \cite{Hogan, Nomura}, and indeed, perhaps even the top quark \cite{Feldstein}, to be outliers.  If this were the case, degenerate neutrino Yukawas might in fact be more representative of the fundamental Yukawa distributions than the hierarchical charged fermion ones.   It might be interesting to see if this picture could be made to work at a quantitative level using some relatively degenerate ansatz for the Yukawa distributions, and putting anthropic constraints on various charged fermions masses.

\section*{Acknowledgments}

We would like to thank Hitoshi Murayama, Taizan Watari and particularly Tsutomu Yanagida for useful discussions and input.
This work was supported by the World Premier International Center Initiative (WPI
Program), MEXT, Japan.

\end{document}